\documentstyle[12pt,aaspp4,psfig]{article}

\newcommand \hii  {H\,{\sc ii}}

\newcommand \ha   {H$\alpha$}
\newcommand \kms  {km~s$^{-1}$}

\newcommand \nii  {[N\,{\sc ii}]}

\begin{document}

\title{Detection of Pre-Shock Dense Circumstellar Material of SN~1978K}

\author{You-Hua Chu\altaffilmark{1}, Adeline Caulet}
\affil{Astronomy Department, University of Illinois, 1002 W. Green Street,
Urbana, IL 61801 \\
Electronic mail: chu@astro.uiuc.edu, caulet@astro.uiuc.edu}

\altaffiltext{1}{Visiting astronomer, Cerro Tololo Inter-American 
Observatory, National Optical Astronomy Observatories, operated by 
the Association of Universities for Research in Astronomy, Inc., 
under a cooperative agreement with the National Science Foundation.}

\author{Marcos J. Montes}
\affil{Naval Research Laboratory, Code 7212, Washington, DC 20375-5320 \\
Electronic mail: montes@rsd.nrl.navy.mil}

\author{Nino Panagia}
\affil{Space Telescope Science Institute, 3700 San Martin Drive, 
Baltimore, MD 21218 \\
Electronic mail: panagia@stsci.edu}

\author{Schuyler D. Van Dyk}
\affil{IPAC/Caltech, Mail Code 100-22, Pasadena, CA 91125 \\
Electronic mail: vandyk@ipac.caltech.edu}

\author{Kurt W. Weiler}
\affil{Naval Research Laboratory, Code 7214, Washington, DC 20375-5320 \\
Electronic mail: weiler@rsd.nrl.navy.mil}

\begin{abstract}

The supernova SN~1978K has been noted for its lack of emission
lines broader than a few thousand \kms\ since its discovery in
1990.  Modeling of the radio spectrum of the peculiar SN 1978K 
indicates the existence of \hii\ absorption along the line 
of sight.  To determine the nature of this absorbing region, 
we have obtained a high-dispersion spectrum of SN~1978K at 
the wavelength range 6530--6610 \AA.  The spectrum shows not 
only the moderately broad \ha\ emission of the supernova 
ejecta but also narrow nebular \ha\ and \nii\ emission.  The high 
\nii$\lambda$6583/\ha\ ratio, 0.8--1.3, suggests that this radio 
absorbing region is a stellar ejecta nebula.  The expansion 
velocity and emission measure of the nebula are consistent 
with those seen in ejecta nebulae of luminous blue variables. 
Previous low-dispersion spectra have detected a strong
\nii$\lambda$5755 line, indicating an electron density of
3--12$\times$10$^5$ cm$^{-3}$.  We argue that this stellar
ejecta nebula is probably part of the pre-shock dense 
circumstellar envelope of SN~1978K.  We further suggest
that SN~1997ab may represent a young version of SN~1978K.

\end{abstract}


%\clearpage

\section{Introduction}

Massive stars lose mass via stellar winds throughout their lifetime.
Stellar winds expand away from the stars and form circumstellar 
envelopes.  As a massive star ends its life in a supernova (SN) 
explosion, the SN ejecta plows through the circumstellar material, 
driving a forward shock into the circumstellar material and a 
reverse shock into the SN ejecta.  Optical emission is generated in 
the ionized SN ejecta, cooled SN ejecta behind the reverse shock, 
shocked circumstellar material, and the ambient ionized circumstellar
material (\markcite{CF94}Chevalier \& Fransson 1994).  These four
regions have different physical conditions and velocity structures.
Consequently, optical luminosities and spectral characteristics
of Type II SNe not only vary rapidly for individual SNe, but also 
differ widely among SNe with different progenitors.

Optical spectra of Type II SNe older than a few years are
characterized by broad hydrogen Balmer lines and oxygen
forbidden lines, with FWHM greater than a few thousand \kms, 
reflecting the rapid expansion of the SN ejecta (e.g., SN~1979C 
and SN~1980K - \markcite{Fe98}Fesen et al.\ 1998; SN~1986E - 
\markcite{CDT95}Cappellaro, Danziger, \& Turatto 1995; 
SN~1987F - \markcite{Fi89}Filippenko 1989; 
SN~1994aj - \markcite{Be98}Benetti et al.\ 1998).  
Some Type II SNe, however, do not seem to show 
such broad emission lines.  The most notable case is SN~1978K.

SN~1978K in NGC\,1313 was discovered in 1990 during a 
spectrophotometric survey of extragalactic \hii\ regions 
(\markcite{RD93}Ryder \& Dopita 1993).  \markcite{Ry93}Ryder 
et al.\ (1993) examined archival optical images of NGC\,1313 
and established that the optical maximum of the supernova 
occurred in 1978, possibly two months before July 31.  
However, the optical spectra of SN~1978K obtained in 
1990--1992 do not show any emission line broader than 600 
\kms\ (Ryder et al.\ 1993; \markcite{CDDV95}Chugai, Danziger, 
\& Della Valle 
1995).  This is in sharp contrast to SN~1980K, which shows 
broad, 6000 \kms\ emission lines in spectra obtained in 1988 
and 1997 (Fesen et al.\ 1998).

SN~1978K is intriguing at radio wavelengths as well.  While
its radio flux shows temporal variations consistent with the
expectation of a typical Type II SN, its radio spectrum
shows a low-frequency turnover that is most plausibly caused
by free-free absorption from an \hii\ region along the line
of sight (Ryder et al.\ 1993).  \markcite{Mo97}Montes, 
Weiler, \& Panagia (1997) re-analyzed the radio observations 
of SN~1978K, and find that the intervening \hii\ region has 
an emission measure EM = $8.5\times10^5 (T_e/10^4 K)^{1.35}$ 
cm$^{-6}$ pc, where $T_e$ is the electron temperature.

To determine the nature of this ``\hii\ region" toward SN~1978K,
we have obtained a high-dispersion echelle spectrum at the 
wavelength range of 6530--6610 \AA.  This spectrum clearly 
resolves the narrow \nii$\lambda\lambda$6548, 6583 lines and a 
narrow \ha\ component from a moderately broad \ha\ component.  
The narrow \ha\ and \nii\ lines must arise from the ``\hii\ 
region", and the broad \ha\ component from the SN ejecta.  
In this paper, we report the echelle observation (\S 2), 
compare our spectrum with previous low-dispersion spectra 
(\S 3), argue that the ``\hii\ region" toward SN~1978K is 
circumstellar, and suggest a feasible explanation for 
SN~1978K's apparent lack of very broad emission lines (\S 4).

\section{High-Dispersion Spectrum of SN~1978K}

We obtained a high-dispersion spectrum of SN~1978K using the 
echelle spectrograph on the 4-m telescope at Cerro Tololo
Inter-American Observatory (CTIO) on 1997 February 27.  The 
spectrograph was used in a long-slit, single-order mode; the 
cross disperser was replaced by a flat mirror and a broad
\ha\ filter (FWHM = 75 \AA) was inserted behind the slit.
The slit width was 250 $\mu$m, or 1\farcs64.  The data were 
recorded with the red long-focus camera and a Tektronix 2048 
$\times$ 2048 CCD.  The pixel size was 0.08 \AA\ pixel$^{-1}$
along the dispersion and 0\farcs26 pixel$^{-1}$ in the spatial 
axis.  The instrumental FWHM was 14$\pm1$ km s$^{-1}$.  The 
data were wavelength-calibrated but not flux-calibrated.

The echelle observation of SN~1978K was made with a 10-min
exposure. SN~1978K and two unrelated \hii\ regions are 
detected. No spatially extended \hii\ features exist at the
position of SN~1978K.  A spectrum extracted from a 5$''$
slit length\footnote{This large slit length is necessary to 
include all the light from SN 1978K, as the telescope was
slightly out of focus for this observation.  When the focus
problem was resolved (2 hours later), SN~1978K had already set.}
centered on SN~1978K 
is presented in Figure 1.  The high-dispersion spectrum
shows three sets of lines with distinct velocity widths.  
The narrowest (unresolved) are the telluric \ha\ and 
OH $\lambda$6553.617 and $\lambda$6577.285 lines 
(\markcite{Os96}Osterbrock et al.\ 1996).  The broadest 
is the \ha\ emission from the supernova ejecta.  It is 
centered at 6572.76$\pm$0.22 \AA, corresponding to a
heliocentric velocity (V$_{hel}$) of 455$\pm$10 \kms;
its FWHM is $\sim$450 \kms\ and FWZI $\sim$1,100 \kms.

The third set of lines consists of the narrow 
\nii$\lambda\lambda$6548, 6583 lines and a narrow \ha\ 
component.  The narrow \ha\ component is superimposed near 
the peak of the broad \ha\ emission of the supernova, hence 
its central velocity, V$_{hel}$ $\sim$ 419 \kms, and FWHM, 
75--100 \kms, are somewhat uncertain.  The \nii$\lambda$6583 
line, at V$_{hel}$ = 419$\pm$5 \kms, shows a line split of 
$\sim$70 \kms; its FWHM is $\sim$125 \kms. The 
\nii$\lambda$6548 line, being weaker, does not show an 
obvious line split; however, its asymmetric line profile 
indicates the presence of a brighter red component and a 
weaker blue component, consistent with those seen in the 
\nii$\lambda$6583 line.  

The narrow \ha\ component and the narrow \nii\ lines most
likely originate from the same emitting region, and will
be referred to as ``nebular" emission.  We have measured 
the nebular \nii$\lambda$6583/\ha\ ratio to be 0.8--1.3.  
The large uncertainty in this ratio is caused by the 
uncertainty in the nebular \ha\ flux, as it is difficult 
to separate the nebular and supernova contributions to the 
observed \ha\ emission.  The possible range of nebular
\nii/\ha\ ratio is derived from the lower and upper limits
of the nebular \ha\ flux, estimated by assuming high and 
low peaks of supernova emission, respectively.

\section{Comparison with Previous Low-Dispersion Spectra}

A relatively low-dispersion spectrum of SN~1978K was obtained on 
1990 Jan 23 by Ryder et al.\ (1993).  That spectrum 
showed an \ha\ line centered at 6570.2$\pm$0.6 \AA\ with a FWHM 
of 563 \kms.  It also detected the \nii$\lambda$6583 line at 
6589.6$\pm$1.0 \AA.  As this spectrum has a resolution of $\sim$5
\AA\ and a pixel size of 1.5 \AA\ pixel$^{-1}$, the \nii\ 
lines are not well resolved from the \ha\ line and consequently 
the velocity and flux measurements might not be very accurate.  
The \nii$\lambda$6583/\ha\ flux ratio, 0.049, derived from this 
low-dispersion spectrum is really the ratio of nebular 
\nii$\lambda$6583 flux to the combined supernova and nebular 
\ha\ flux.  The \nii$\lambda$5755 line is also detected and the 
\nii$\lambda$5755/\ha\ flux ratio is 0.025.

Another low-dispersion spectrum of SN~1978K was obtained on
1992 October 22 by \markcite{CDD95}Chugai et al.\ (1995).  
The resolution of this spectrum is 10 \AA.  
Thus the redshifts and widths of spectral lines cannot be 
reliably determined.  The \nii$\lambda$6583/\ha\ flux ratio is 
0.072, and the \nii$\lambda$5755/\ha\ flux ratio is 0.016.

Using our echelle spectrum, we have measured the ratio of 
nebular \nii$\lambda$6583 flux to the combined supernova and 
nebular \ha\ flux to be 0.06.  This is different from the
previous measurements, 0.049 and 0.072.
While our measurement should be more accurate because of our
higher spectral resolution, the supernova \ha\ flux might have 
varied from 1990 to 1997 (Chugai et al.\ 1995).  It is not 
clear whether the \nii\ flux itself has changed.

Nebular lines toward SN~1978K are also detected in the UV 
spectra of SN~1978K obtained with the Faint Object Spectrograph
on board the Hubble Space Telescope on 1994 September 26 and 
1996 September 22--23 (Schlegel et al.\ 1998).  The Ly$\alpha$
line and the blended [Ne~{\sc iv}]$\lambda\lambda$2421, 2424 
doublet are detected.  Both lines have FWHMs comparable to the 
instrumental resolution, 7 \AA, corresponding to 1727 \kms\ at 
Ly$\alpha$ and 866 \kms\ at [Ne~{\sc iv}].  These [Ne~{\sc iv}]
lines have critical densities of 8$\times$10$^4$ and 
2.5$\times$10$^5$ cm$^{-3}$, respectively (\markcite{Zh88}Zheng 
1988); therefore, these [Ne~{\sc iv}] lines must originate from 
the nebula.  The Ly$\alpha$ line emission, like the H$\alpha$ 
emission, contains both the supernova ejecta and nebular 
components.

\section{Discussion}

\subsection{Origin of the Narrow \ha\ and \nii\ Lines}

The most intriguing features detected in our high-dispersion 
spectrum of SN~1978K are the narrow nebular \ha\ and \nii\ 
lines, which are presumably emitted by the ``\hii\ region 
along the line of sight" implied by the radio spectrum of 
SN~1978K (Ryder et al.\ 1993).  However, as we argue below, 
the \nii\ line strengths suggest that this ``\hii\ region" 
is circumstellar, rather than interstellar.

The nebular \nii$\lambda$6583/\ha\ line ratio, 0.8--1.3, 
is unusually high for normal interstellar \hii\ regions in 
a spiral galaxy.  For example, \hii\ regions in M101 have 
\nii$\lambda$6583/\ha\ ratios $\le$0.3 
(\markcite{KG96}Kennicutt \& Garnett 1996).  SN~1978K is at 
the outskirts of NGC\,1313, where abundances are expected to
be low and the \hii\ excitation is expected to be high.  If 
the nebular \ha\ and \nii\ lines toward SN~1978K originate 
in an interstellar \hii\ region, we would expect the 
\nii$\lambda$6583/\ha\ ratio to be $\sim$0.1 or lower.  
A low interstellar \nii/\ha\ ratio is confirmed by the bright 
\hii\ region detected along the slit at $\sim90''$ east of 
SN~1978K.  This \hii\ region is brighter than the nebula 
toward SN~1978K in the \ha\ line, but its \nii$\lambda$6583
line is not detected.  We may rule out an interstellar 
\hii\ region explanation for the narrow nebular lines seen 
in SN~1978K.

The high \nii$\lambda$6583/\ha\ ratio may be caused by a high 
electron temperature or a high nitrogen abundance.  These 
conditions can be easily provided by SN~1978K and its progenitor.
If the nebula was ionized by the UV flash of SN~1978K, the 
electron temperature may be higher than that of a normal \hii\
region, as in the case of SN~1987A's outer rings 
(\markcite{Pa96}Panagia et al.\ 1996).  However, the 
\nii$\lambda$6583 line 
intensity increases by only a factor of 2 for an electron 
temperature increase from 10,000 K to 15,000 K.  This increase
cannot explain fully the observed high \nii/\ha\ ratio.  
A higher nitrogen abundance is needed.  An elevated nitrogen 
abundance is characteristic of ejecta nebulae around evolved 
massive stars, such as luminous blue variables (LBVs) and 
Wolf-Rayet (WR) stars; the \nii$\lambda$6583/\ha\ ratios 
of these ejecta nebulae are frequently observed to be $\sim$1 
(\markcite{Es92}Esteban et al.\ 1992; \markcite{Sm98}Smith 
et al.\ 1998).  Therefore, the most reasonable origin of the
nebular emission lines toward SN~1978K would be a 
circumstellar ejecta nebula.  The observed high \nii/\ha\ 
ratio may be caused by the combination a high nitrogen 
abundance and a high electron temperature.

SN~1978K's circumstellar ejecta nebula has a very high density,
as strong \nii$\lambda$5755 line is observed in SN~1978K's 
spectrum.  The \nii\ ($\lambda$6548+$\lambda$6583)/$\lambda$5755
ratio is measured to be 2.55 by Ryder et al.\ (1993), and 6.0 
by Chugai et al.\ (1995), indicating that collisional 
de-excitation is significant for the $^1D_2$ level of N$^+$.
If we assume an electron temperature of 1--1.5$\times10^4$ K, 
the observed \nii\ line ratios imply electron densities of 
3--12$\times10^5$ cm$^{-3}$.  

The circumstellar ejecta nebula of SN~1978K can be compared 
to those observed around LBVs and WR stars.  The density
of SN~1978K's nebula is higher than those of WR nebulae,
but within the range for LBV nebulae (\markcite{St89}Stahl 
1989; \markcite{Es92}Esteban et al.\ 1992).  We adopt the
emission measure EM = $8.5\times10^5 (T_e/10^4 K)^{1.35}$ 
cm$^{-6}$ pc determined from the radio observations (Montes 
et al.\ 1997) for SN~1978K's nebula.  This emission measure 
is much higher than those observed in ejecta nebulae around 
WR stars, typically a few $\times$10$^2$ to 10$^3$ cm$^{-6}$ 
pc (\markcite{Es92}Esteban et al.\ 1992; 
\markcite{EV92}Esteban \& V\'{\i}lchez 1992), but 
lies toward the high end of the range typically seen in LBV 
nebulae, a few $\times$10$^3$ to 10$^5$ cm$^{-6}$ pc 
(\markcite{Hu94}Hutsem\'ekers 1994; 
\markcite{Sm98}Smith et al.\ 1998).  
Finally, the \ha\ and \nii\ velocity profiles seen in our 
SN~1978K spectrum suggest an expansion 
velocity\footnote{The expansion velocity implied by the 
line split in the \nii\ line is $>$35 \kms.  The expansion 
velocity can also be approximated by the HWHM of the \ha\ 
and \nii\ lines, 40--55 \kms.}
of 40--55 \kms, which is lower than those of most ejecta 
nebulae around WR stars but is within the range for LBV nebulae 
(\markcite{No95}Nota et al.\ 1995; 
\markcite{Ch98}Chu, Weis, \& Garnett 1999).  It is thus 
likely that the observed nebula toward SN~1978K was ejected 
by the progenitor during a LBV phase before the SN explosion.

This ejecta nebula could be either part of the circumstellar
envelope that the SN ejecta expands into, or a shell that is 
detached from the circumstellar envelope.  We will demonstrate
below that the latter is unlikely.  If the ejecta nebula is a
detached shell, the observed emission measure and density 
imply that the shell thickness is only
4$\times$10$^{-5}$ to 4$\times$10$^{-6}$ pc.
The thickness of a detached, dense shell will be broadened 
by diffusion, and may be crudely approximated by
$(c/V_{exp})R$, where $c$ is the isothermal sound velocity, 
$V_{exp}$ is the expansion velocity, and $R$ is the radius.
We find that the radius of SN~1978K's free-expanding ejecta 
shell would have to be no greater than $\sim2\times10^{-4}$ 
pc, which is smaller than the expected radius of the SN 
ejecta.  This is impossible.  Therefore, we conclude that 
the narrow \ha\ and \nii\ lines must originate in the 
pre-shock, ionized circumstellar envelope of SN~1978K.

The narrow nebular lines from the pre-shock, ionized 
circumstellar envelope of SN~1978K are not unique among 
SNe.  The high-dispersion spectrum of SN~1997ab shows 
narrow P-Cygni \ha\ and narrow \nii$\lambda$6583 lines,
and the FWZI of the P-Cygni \ha\ line, 180 \kms, is 
comparable to that of SN~1978K's \ha\ line 
(\markcite{Sa98}Salamanca et al.\ 1998).  
The P-Cygni profile of SN~1997ab's narrow 
\ha\ line indicates a high density, $\ge 10^7$ cm$^{-3}$.
This density exceeds the  critical density of the $^1D_2$
level of N$^+$, and causes a weak \nii$\lambda$6584 line
(see Figure 1 of Salamanca et al.\ 1998).
If SN~1997ab's circumstellar material is 
nitrogen-rich like that of SN~1978K, we predict that its 
\nii$\lambda$5755 line is strong and should be detectable
as well.  SN~1997ab is very likely a younger version of
SN~1978K, and SN~1978K's nebular \ha\ line may have 
exhibited a P-Cygni profile in 1979-1980.

\subsection{SN Evolution in a Very Dense Circumstellar Envelope}

The most notable SN characteristic of SN~1978K is its apparent
lack of very broad (a few thousand \kms) emission lines.  
Adopting canonical expansion velocities and sizes for SN~1978K,
Ryder et al.\ (1993) has derived a mass of $>$80 M$_\odot$ for 
the circumstellar envelope.  This mass is too large to reconcile 
with the current understanding of massive stellar evolution.
To lower the circumstellar mass, Chugai et al.\ (1995) propose 
that the circumstellar envelope is clumpy.

We consider that the large size $\sim$0.1 pc adopted by Ryder
et al.\ (1993) is over-estimated and inconsistent with the
expansion velocity implied by our observed \ha\ FWHM of 450 
\kms.  There is no need to assume an unseen, larger expansion 
velocity.  We suggest that the small expansion velocity of 
SN~1978K is caused by the dense circumstellar envelope, which 
has quickly decelerated the expansion of SN ejecta.  If optical 
spectra had been obtained immediately after the SN explosion
in 1978, very broad emission lines would have been detected.  

Rapid deceleration of SN ejecta has been observed in two other
SNe, SN~1986J and SN~1997ab.  SN~1986J has been noted to have 
very similar spectral properties as SN~1978K\footnote{We have
examined a large number of SN spectra reported in the literature.
SN~1986J appears to be the only SN besides SN~1978K that shows 
strong \nii$\lambda$5755 line, indicating a very high density 
and possibly an enhanced nitrogen abundance.}.
SN~1986J probably exploded four years before its initial 
discovery in 1986 (\markcite{Ru87}Rupen et al.\ 1987; 
\markcite{Ch87}Chevalier 1987).  
Its optical spectra obtained soon after the discovery show 
narrow hydrogen Balmer lines and nitrogen forbidden lines, 
indicating an expansion velocity $<$ 600 \kms\ 
(\markcite{Le91}Leibundgut et al.\ 1991).  
SN~1997ab is the only other SN for which narrow nebular emission
lines from the dense circumstellar envelope have been 
unambiguously resolved and detected.  SN~1997ab's light curve 
peaked in 1996; the FWHM of its \ha\ line decreased rapidly 
from 2500 \kms\ on 1997 March 2 to 1800 \kms\ on 1997 May 30 
(\markcite{Ha97}Hagen et al.\ 1997; Salamanca et al.\ 1998).

Clearly, SN~1978K, SN~1986J, and SN~1997ab all possess very
dense circumstellar envelopes, and we may expect them to evolve
similarly.  The expansion of SN~1978K might have slowed 
down to below 1000 \kms\ within the first $\sim$2 years after
the explosion, and the SN ejecta could not have reached a 
radius greater than $\sim$0.02 pc in 1990.  A factor of 5
reduction in the radius would lower Ryder et al.'s (1993)
estimate of mass to a reasonable value, and the hypothesis of
a clumpy circumstellar envelope will no longer be necessary.

\subsection{Future Work}

Previous spectrophotometric observations of SNe were rarely 
made with spectral resolutions better than 2 \AA.  
Our echelle observation 
of SN~1978K has demonstrated that high-dispersion spectroscopy
is powerful in resolving pre-shock, ionized circumstellar 
material.  A high-dispersion spectroscopic survey of young 
SNe in nearby galaxies may detect more circumstellar envelopes
and even detached ejecta ring nebulae, such as the rings around
SN~1987A (\markcite{Bu95}Burrows et al.\ 1995).
The density and velocity structures of these circumstellar
envelopes would shed light on the mass loss history as well
as physical properties of the massive progenitors.

Our spectrum of SN~1978K unfortunately covers only the \nii\ 
and \ha\ lines.  In order to measure the density, temperature, 
and abundances of the circumstellar material, it is necessary 
to obtain high-dispersion spectra covering a large wavelength 
range.  It is also important to monitor the spectral changes 
indicative of density changes in the circumstellar envelope.
A large change at all wavelengths is expected when the SN
ejecta expands past the outer edge of the circumstellar envelope.

\acknowledgments 
We would like to thank the referee for useful suggestions to 
improve this paper.  YHC acknowledges the support of NASA LTSA 
grant NAG 5-3246. MJM and KWW wish to thank the Office of Naval 
Research (ONR) for the 6.1 funding supporting this research.

\clearpage
%\vskip 7cm
\begin{center} {\large \bf Figure Captions} \end{center}

\begin{figure}[tbh]
\centerline{\psfig{file=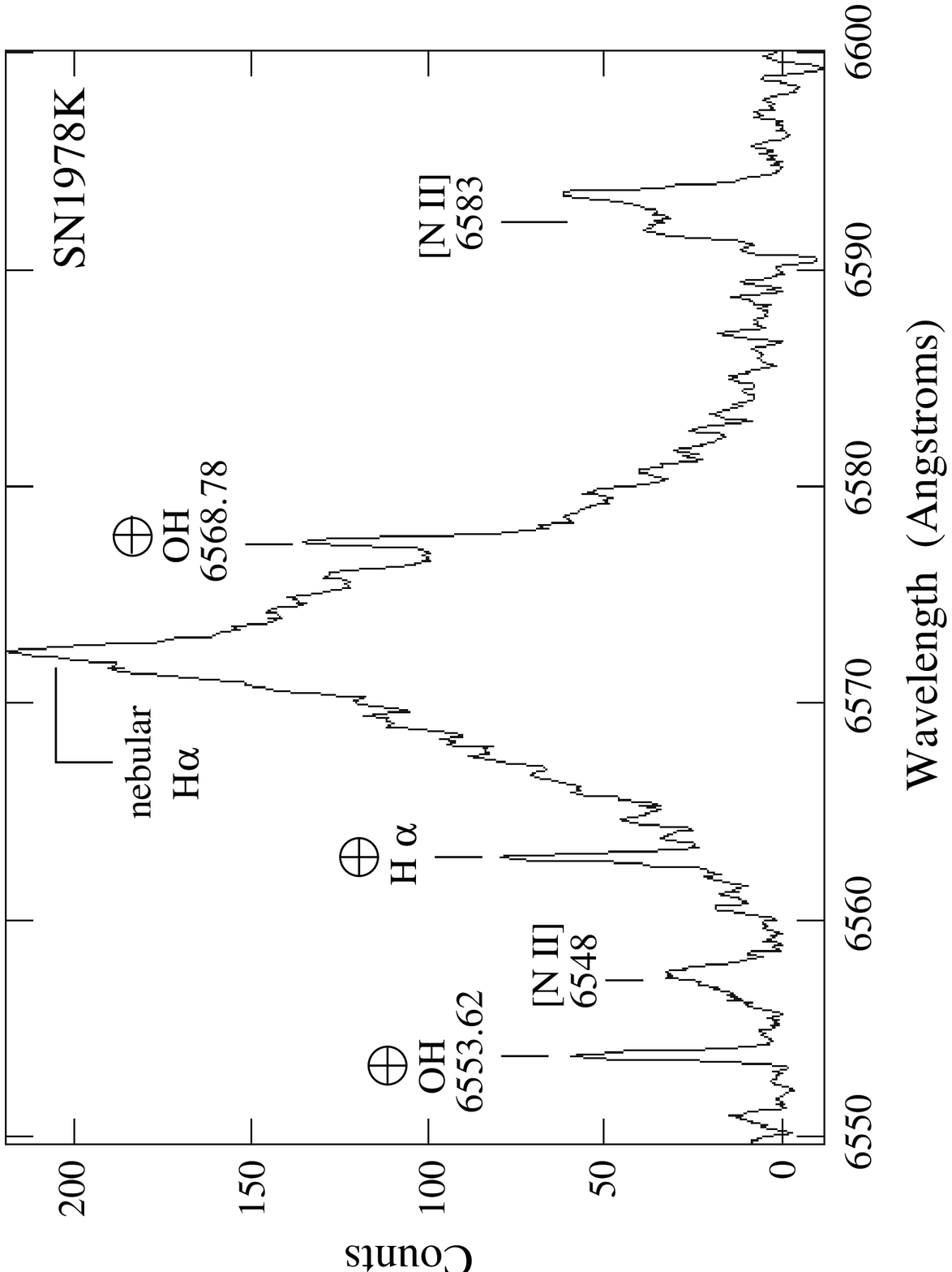}}
\end{figure}

\figcaption{High-dispersion spectrum of SN~1978K taken with the
echelle spectrograph on the CTIO 4 m telescope.  The spectrum has
been smoothed with a boxcar of 5 pixels, or 0.4 \AA.  The telluric 
lines are indicated by the $\oplus$ symbol.}

\end{document}